\pdfoutput=1
\documentclass{article}
\usepackage{spconfa4,amsmath,graphicx}
\usepackage{todonotes}
\usepackage[normalem]{ulem}

\usepackage{comment}
\usepackage{bm}
\usepackage{booktabs}
\usepackage{svg}
\usepackage{scalerel}
\usepackage{etoolbox}
\usepackage{multirow}

\BeforeBeginEnvironment{figure}{\vskip-2ex}
\AfterEndEnvironment{figure}{\vskip-10ex}

\title{Deep neural network based speech separation optimizing  an objective estimator of intelligibility for low latency applications \vspace{0mm}} 
\name{Gaurav Naithani\textsuperscript{1} \thanks{The authors wish to thank CSC-IT Centre of Science Ltd., Finland,  for providing computational resources used in  experiments reported in this paper.},  Joonas Nikunen \textsuperscript{1}, Lars Bramsl\o{}w\textsuperscript{2}, Tuomas Virtanen\textsuperscript{1} \vspace{-3mm}}
\address{ \textsuperscript{1}Tampere University of Technology, Laboratory of Signal Processing, Tampere, Finland\\
\normalsize Email:\{gaurav.naithani, joonas.nikunen,  tuomas.virtanen\}@tut.fi  \vspace{0mm}\\
\textsuperscript{2}Eriksholm Research Centre, Oticon A/S, Snekkersten, Denmark\\
\normalsize Email:\{labw \}@eriksholm.com\\
\vspace{-8mm}}
\begin{document}
%
\maketitle
\begin{abstract}
Mean square error (MSE) has been the preferred choice as loss function in the current deep neural network (DNN) based speech separation techniques. In this paper, we propose a new cost function with the aim of optimizing the extended short time objective intelligibility (ESTOI) measure.  We focus on applications where low algorithmic latency \mbox{($\leq 10$ ms)} is important. We use long short-term memory networks (LSTM) and evaluate our proposed approach on  four sets of two-speaker mixtures from extended Danish hearing in noise (HINT) dataset. We show that the proposed loss function can offer improved or at par objective intelligibility (in terms of ESTOI) compared to an MSE optimized baseline while resulting in lower objective separation performance (in terms of the source to distortion ratio (SDR)). We then proceed to propose an approach where the network is first initialized with weights optimized for MSE criterion and then trained with the proposed ESTOI loss criterion. This approach mitigates some of the losses in objective separation performance while preserving the gains in objective intelligibility. 


\end{abstract}

\begin{keywords}
Deep neural networks, Speech separation, Speech intelligibility, Low latency.
\end{keywords}
\section{Introduction}\label{sec:intro}
\vspace{-1ex}
Monaural speech separation is the problem of separating a target speech signal from an acoustic mixture consisting of other highly non-stationary signals, e.g., competing speech signals. Traditionally, model based approaches like hidden Markov models (e.g., in \cite{roweis2001one} ) and non-negative matrix factorization (e.g., in \cite{virtanen2007monaural, schmidt2006single}) have been used to address it. In recent years, however, purely data driven discriminative approaches like deep neural networks (DNNs) (e.g., in \cite{huang2014deep, erdogan2015phase}) have achieved great success.  


In this paper, we focus on objective intelligibility performance of DNN-based speech separation. Moreover, our approach concerns with maintaining a low algorithmic processing latency (e.g., in \cite{tasnet2018, naithani2017low, naithani2016low}) which is particularly critical for applications like hearing aids \cite{bramslow2010preferred} and cochlear implants \cite{hidalgo2012low}. Notably for hearing aids, according to  Agnew~{\textit{et~al}.~\cite{agnew2000hearing}, delays as low as  3 to \mbox{5 ms} were found to be noticeable and anything longer than \mbox{10 ms} was deemed objectionable to hearing impaired listeners due to potential comb filter coloration or echo from the combination of direct and delayed sound in open hearing aid fittings. 

DNN-based speech separation approaches have generally been using the traditional mean square error (MSE) loss function (e.g., in \cite{huang2014deep, xu2015regression}) between the predicted and target spectrum or time-frequency masks. MSE loss is intuitively sub optimal in the sense that it treats all frequency components of the signal equally which deviates from what has been suggested from the studies of human auditory system \cite{loizou2007speech}. Hence it makes sense to use perceptually motivated cost functions instead. There have been some attempts towards this goal in  the context of speech separation/enhancement. For example, an altered version of traditional MSE was used in \cite{shivakumar2016perception}, a weighted MSE approach based on absolute threshold of hearing and masking properties of human auditory system was  employed in \cite{kumar2016speech}, and a cost function inspired from short-time objective intelligibility (STOI) \cite{taal2010short} measure was used in \cite{kolbaek2018monaural}. Another notable work comparing different cost functions, e.g., Kullback-Leibler divergence, Itakura-Saito divergence, and, MSE, was reported in \cite{nugraha2016multichannel}.

\begin{figure*}[h!]
\centering
\includegraphics[scale=0.44]{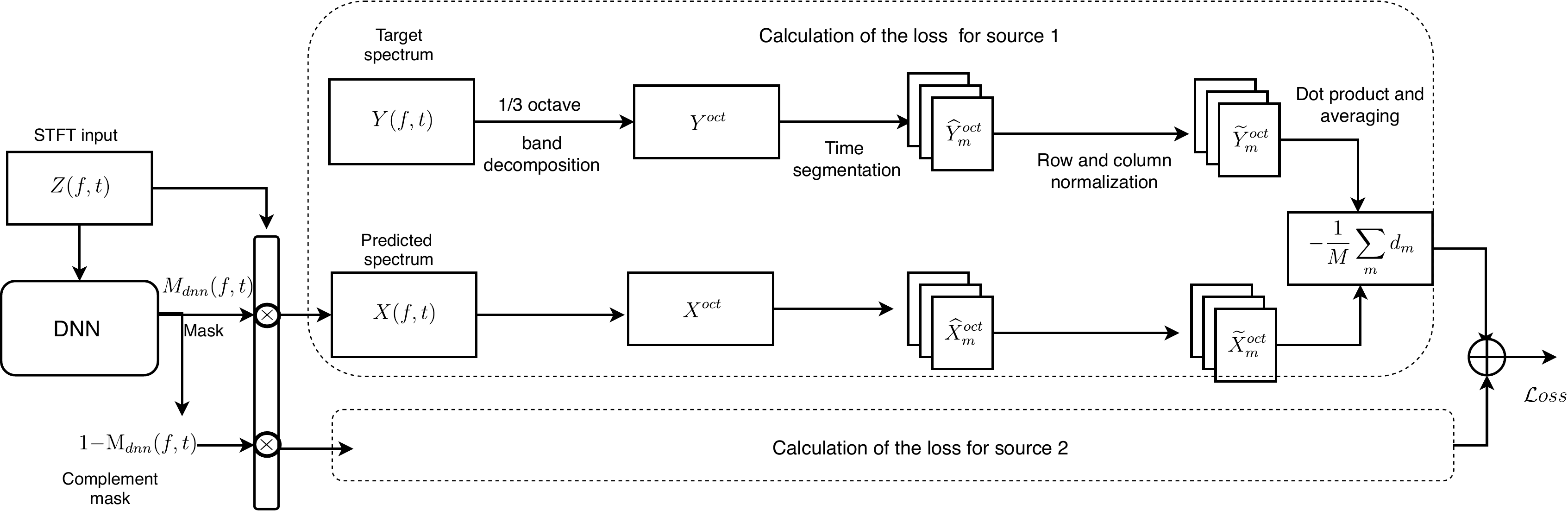}
\vspace{-1ex}
\caption{\small The proposed DNN scheme with objective function incorporating extended short time objective intelligibility (ESTOI) measure. Loss calculation corresponding \textit{source 2} is identical to the one shown for \textit{source 1}. }
\label{fig: estoi_loss}
\end{figure*}

We propose a cost function based on the extended short time objective intelligibility (ESTOI) metric \cite{jensen2016algorithm}. ESTOI extends the widely used short time objective intelligibility (STOI) metric \cite{taal2010short} and is postulated to be a  better predictor of subjective intelligibility  when the interfering signal is modulated \cite{jensen2016algorithm}, e.g.,  competing speech,  and hence is better suited  as an optimization objective for our purpose.  We optimize for ESTOI using a sequence based loss using a long short-term memory network unlike \cite{kolbaek2018monaural} where feedforward DNNs were used for STOI optimization. Moreover, we  use a single network which jointly optimizes for all one-third octave bands unlike \cite{kolbaek2018monaural} where multiple networks were used.


In direct optimization with ESTOI loss, we report an improvement of 0.03 (\textit{p} $<0.05$, Wilcoxon signed-rank test ) averaged over four speaker pairs in terms of ESTOI metric as compared to a baseline DNN trained on MSE cost. We also observe that this optimization degrades the separation performance in terms of source to distortion ratio (SDR) \cite{vincent2006performance} on an average by 0.6 dB. We then propose a pretraining strategy where the DNN is first trained with MSE as objective and continue the training after model convergence (in terms of validation MSE) with the ESTOI loss. The proposed approach mitigates the degradation in SDR to an average of 0.2 dB  while offering a better or at par objective intelligibility performance compared to the baseline. 

\section{Proposed Cost function} \label{sec: cost}
\vspace{-1ex}
In this paper we propose a sequence based loss that approximates the calculation of ESTOI estimator. The  magnitude spectrogram of the mixture, $Z(t,f)$, is computed,  $f$ and $t$ being frequency and time indices, respectively, and fed to an LSTM. Each sequence consists  of $T$ successive STFT frames. The proposed loss is computed between estimated and target spectra of the two sources. Both STOI and ESTOI measures utilize one-third octave band processing to mimic frequency selectivity of cochlear processing in human ear. Moreover, ESTOI measure is computed on short time analysis segments of 384 ms in order to include  temporal modulation frequencies which are critical for speech intelligibility \cite{taal2011algorithm}. We  keep these design choices consistent in our cost calculation with the only difference from the original ESTOI computation being the inclusion of frequency range  up to 8 kHz.  For the sake of simplicity we explain loss calculation between the estimated spectrum, $X(f,t)$ and  target spectrum $Y(f, t)$  corresponding to one source only and the process is identical for the second source. It entails the following steps:




\begin{enumerate}
\vspace{-1ex}
\item \textit{Band decomposition}:  $Y(f, t)$ is processed to give one-third octave band decomposed version $Y^{oct}(j,t)$ as,
\vspace{-1ex}
\begin{equation}
\begin{aligned}
Y^{oct}(j,t) = \sqrt{\sum_{f=f1(j)}^{f=f2(j)}|Y(f, t)|^2}, \quad j=1,..,J, 
\end{aligned}
\end{equation}
where $f_1(j)$ and $f_2(j)$ denote the frequency boundaries of $j^{th}$ one-third octave band. $J$ is the number of one-third octave bands. Similarly, $X^{oct}(j,t)$ is one-third octave band decomposed version obtained  from $X(f,t)$.

\item \textit{Time segmentation}: Now $Y^{oct}(j,t)$ (and $X^{oct}(j,t)$) is segmented into $T-N+1$  time-segments, where $T$ is the number of STFT frames in $Y^{oct}$, $N$ is ESTOI context window for the calculation of intermediate intelligibility measures. Hence $m^{th}$ time-segment for $Y^{oct}$ is a $J \times N$ matrix given by, 
\vspace{-1ex}
\begin{equation*}
\widehat{Y}^{oct}_m=\begin{bmatrix}
Y^{oct}(1, m-N+1) & \cdots & Y^{oct}(1,m)\\
\vdots & \vdots  & \vdots \\
Y^{oct}(J, m-N+1) & \cdots & Y^{oct}(J, m)\\
\end{bmatrix} \quad .
\end{equation*}
Similarly, $\widehat{X}^{oct}_m$ is the $m^{th}$ time segment corresponding to band decomposed spectrogram $X^{oct}$ for intermediate ESTOI calculation.

\item \textit{Normalization}: Each of the above segments is then mean  and variance normalized first along the rows (temporal normalization) such that each row of resulting matrix  is zero mean and unit norm. It is followed by normalization along columns (spectral normalization) yielding  a matrix \mbox{$\widetilde{Y}^{oct}_m = \left[\tilde{\textbf{y}}_{1,m} \cdot \cdot \cdot \tilde{\textbf{y}}_{N,m} \right]$},  each column of which is a unit norm and zero mean vector.

\item \textit{Dot product and averaging}: The intermediate intelligibility index corresponding to time segment $m$ is simply the dot product of columns of $\widetilde{Y}^{oct}_m$ and $\widetilde{X}^{oct}_m$ given by,
\vspace{-1ex}
\begin{equation}
\vspace{-1ex}
d_m=\frac{1}{N}\sum_{n=1}^{N} \tilde{\textbf{x}}_{n,m}^{\top} \tilde{\textbf{y}}_{n,m} \,.
\end{equation}
Where $\top$ denotes the transpose operation. The \mbox{ESTOI} metric corresponding to the sequence is then calculated by averaging the $M$ intermediate measures, i.e. $d_{final}= \frac{1}{M}\sum_m d_m$. The cost aims to maximize this metric which can be achieved by minimizing negative of the ESTOI metric for the sequence, i.e, minimizing $-d_{final}$. 
\end{enumerate}
\vspace{-1ex}
For simplicity, Figure \ref{fig: estoi_loss} depicts the process of computation of the loss function for \textit{source 1} only. Similar loss calculation is done for \textit{source 2} and final loss is mean of the two losses. All the operations described above are differentiable. Libraries Keras \cite{chollet2016keras} and Theano \cite{theano} are used for training which performs automatic differentiation and gradient backpropagation.

We consider  a long short-term memory network (LSTM) \cite{hochreiter1997long} as the baseline DNN topology. There are three cases under investigation here: \textit{a)}  MSE objective,   we will denote it as MSE-DNN, \textit{b)} The proposed ESTOI objective, we will denote it as ESTOI-DNN, and, \textit{c)} Training with the proposed objective but instead of training from the scratch we use weights from the first  case as initial weights, we will denote it as \mbox{MSE-ESTOI-DNN}. Please see  \ref{subsec:results} for the discussion on the motivation for \textit{Case c}.

\section{Implicit Time-frequency masking } \label{sec: baseline_dnn}  
\vspace{-1ex}
In this work, we use masking based source separation paradigm (e.g., used in \cite{huang2014deep, weninger2014discriminatively, zhao2016dnn}) where a DNN is used to predict a time-frequency mask corresponding to a target speaker. The ESTOI computation however is done for estimated and reference source spectra. We adopt an implicit mask prediction scheme in the sense that DNN is being optimized to output mask such that when mask is applied element-wise to the mixture, the resulting source spectrum minimizes the loss calculated in the spectrum domain. A similar scheme was used in \cite{huang2014deep},   and in \cite{mimilakis2017recurrent} in the form of skip filtering connections. The predicted spectrum  for the target \textit{source 1}, $X(f,t)$ gets computed from predicted mask $M_{dnn}(f,t)$ as, 
\vspace{-1ex}
\begin{equation} \label{eq:mask_mul1}
\vspace{-1ex}
X(f,t)= M_{dnn}(f,t) \odot Z(f,t),
\end{equation}
where $\odot$ denotes the Hadamard product. The mask corresponding to the other speaker is defined to be $1- M_{dnn}(f, t)$ and hence the corresponding predicted spectrum is, 
\vspace{-1ex}
\begin{equation} \label{eq:mask_mul2}
\vspace{-1ex}
X^2(f,t)= (1-M_{dnn}(f,t)) \odot Z(f,t).
\end{equation}

\noindent We incorporate above two masking operations as a deterministic layer at the network output and jointly estimate the output spectra corresponding to the two sources similar to \cite{huang2014deep}. This also enforces the condition of the sum of two masks being equal to $1$.

\begin{figure*}[h!]
\centering
\includegraphics[trim = 20mm 6mm 20mm 8mm, clip,height=0.6\columnwidth, width=2\columnwidth]{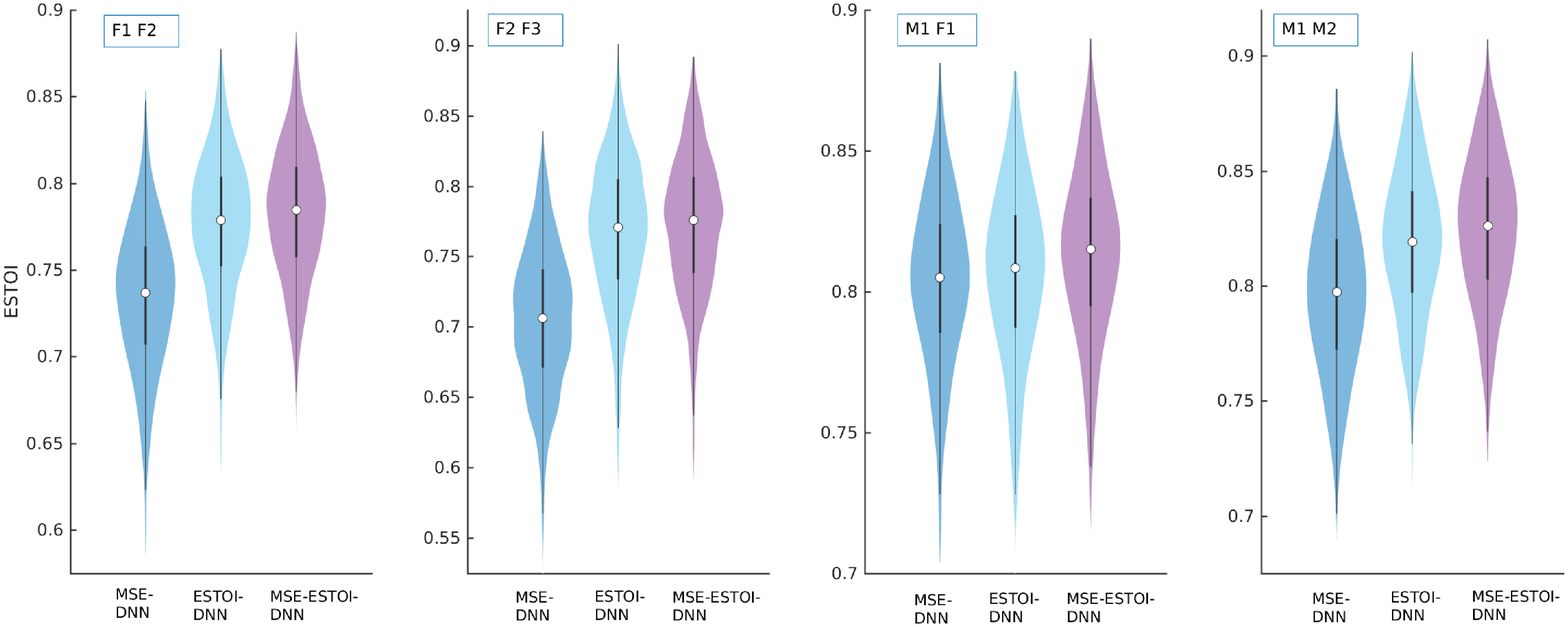} 
\vspace{-1ex}
\caption{ \small Comparison of ESTOI metric values for the three DNN configurations: MSE-DNN, ESTOI-DNN, and, \mbox{MSE-ESTOI-DNN}, corresponding to the four evaluation speaker pairs.}
\label{fig: estoi_LSTM}
\end{figure*}


\section{Evaluation} \label{sec: eval}
\vspace{-1ex}
This section describes the acoustic material used in the experiments, metrics used to evaluate the separation and intelligibility performance of the proposed system and finally the results obtained.

\subsection{Acoustic Material and data generation} 
\vspace{-1ex}
The Danish hearing in noise test (HINT) dataset is used for experiments reported in this paper. It is an  extended version of Danish HINT dataset  \cite{Nielsen2011} and consists of three male and three female speakers. Each speaker has 13 lists, each consisting of  20 five word  sentences of natural speech. The native sampling rate is 44.1 kHz which is downsampled to 16 kHz before processing. Four speaker pairs: F1 and F2, F2 and F3, M1 and F1; and M1 and M2, are used for the evaluation. A separate network is trained for each of these. Eight lists are used (L6 to L13) for training, two lists (L4, L5) for validation, and two lists (L1, L2)  for testing. Total duration of audio for training and validation is approximately 7 minutes. 

STFT spectra are used as DNN input features with analysis window of 128 samples (8 ms) and 50 \% frame overlap, resulting in 8 ms algorithmic latency.  For generating the training data, all available audio signals are concatenated in the time domain corresponding to each speaker. STFT features are then extracted. As the available training material is quite low, a data augmentation scheme is used to increase the  amount of training data. It involves circularly offsetting one speaker spectrogram with respect to the other and adding them to generate mixture spectrogram. Note that the summation here is in complex domain. In this work, we use 30 shifts of temporal length $n\times\frac{T_s}{30}$, where $n=1,..,30$ and $T_s$ is the number of STFT frames in the longer of the two training spectrograms. It effectively increases the amount of training  data by a factor of 30 to around 2.6 hours. 

\subsection{Metrics}
\vspace{-1ex}
BSS-EVAL toolbox \cite{vincent2006performance} is used for objective evaluation of separation performance and ESTOI metric was used for evaluation of speech intelligibility. For the former, we report SDRs. Source to interference ratio (SIR) and source to artifact ratio (SAR) are also reported for completeness. In addition, we also report STOI values as well as it is more widely reported being an older measure.

\subsection{Experimental design}
\vspace{-1ex}

The design choices for ESTOI computation in the proposed loss function are kept inline with the standard ESTOI computation, i.e., ESTOI context for intermediate correlation measures $d_m$ is 384 ms and the centre frequency of lowest one-third octave band is set at 150 Hz. The frequency range used however is up to 8 kHz. The LSTM network uses three hidden layers, each having 512 hidden neurons and a time-distributed feedforward dense layer as the output.  The sequence length used here is 256 STFT frames ($1.02 s$) to have enough time context for several intermediate ESTOI calculation segments. LSTM cells used in the recurrent layers here are standard as described in \cite{gers1999learning}.  The Adam  optimizer is used with default parameters as recommended in \cite{kingma2014adam}. A patience value of 30 epochs is used which means the training is stopped when the error on validation data does not got down for 30 consecutive epochs. For audio processing and feature extraction Librosa \cite{librosa} library is used. The experiments are conducted for five initialization seeds and averaged to get the final results reported here.

\subsection{Results}\label{subsec:results}
\vspace{-1ex}

\begin{table}[b!]
\footnotesize
\centering
\caption{\small The mean objective evaluation metrics for the three DNN configurations: MSE-DNN, ESTOI-DNN, and, \mbox{MSE-ESTOI-DNN}, corresponding to the four speaker pairs.}
\label{tab:results}
\vspace{-1ex}
\begin{tabular}{l@{\hspace{1ex}}l@{\hspace{0ex}}ccccc}
\toprule
Speaker          & \hspace{1mm} DNN     & ESTOI     & STOI   & SDR     & SIR    & SAR   \\ 
\midrule
\multirow{3}{*}{F1 F2} & MSE-DNN        & 0.73      &  0.83  & 7.3     & 11.0  & 10.2  \\
                       & ESTOI-DNN      & 0.78      &  0.86  & 6.8     & 10.5  & 9.9  \\
                       & MSE-ESTOI-DNN  & 0.78      &  0.86  & 7.1     & 10.9  &  10.0 \\           
\midrule                       
\multirow{3}{*}{F2 F3} & MSE-DNN        & 0.70      &  0.81  & 5.8     & 9.1   & 9.4  \\
                       & ESTOI-DNN      & 0.77      &  0.86  & 5.8     & 9.2   &  9.2  \\
                       & MSE-ESTOI-DNN  & 0.77      &  0.86  & 6.0     & 9.6   &  9.2 \\                         
\midrule
\multirow{3}{*}{M1 F1} & MSE-DNN        & 0.80      &  0.90  & 8.6     & 12.4  & 11.3 \\
                       & ESTOI-DNN      & 0.80      &  0.90  & 7.0     & 10.6  &  10.2  \\
                       & MSE-ESTOI-DNN  & 0.81      & 0.90   & 8.1     & 11.8  &  10.9 \\                                     
\midrule
\multirow{3}{*}{M1 M2} & MSE-DNN        & 0.80      &  0.90  & 7.9     & 12.1  & 10.4 \\
                       & ESTOI-DNN      & 0.82      &  0.91  & 6.8     & 10.4  &  10.0 \\
                       & MSE-ESTOI-DNN  & 0.82      &  0.91  & 7.5     & 11.5  &  10.3  \\                                  
\bottomrule                       
\end{tabular}
\end{table}

For the evaluation, list L1 for first speaker of the pair and list L2 for the second speaker are used. Each list  consists of 20 sentences and hence we have 400 test mixtures for each speaker pair. The Table \ref{tab:results} shows the mean objective evaluation metrics for the four speaker pairs.  Moreover, Figure \ref{fig: estoi_LSTM} depicts the violin plots of ESTOI values for the four speaker pairs which shows the distribution of metric values  in addition to the embedded boxplot with median and interquartile range. With ESTOI-DNN, an average improvement of 0.03 in terms of ESTOI metric is observed. An important consequence of ESTOI optimization is poorer separation performance in terms of SDR in all speaker pairs except F2F3 as compared to MSE-DNN baseline. On an average, a degradation of 0.6 dB is observed. Loss functions aiming to improve objective intelligibility may result in decrease in other signal energy based separation metrics, such as SDR. However, the aim is to also maintain on par objective separation criteria to ensure good subjective quality of the separated signal. This observation partly motivates \textit{Case c}. Hence  instead of training models from scratch we use weights of MSE-DNN as initial weights and train for ESTOI objective. DNNs trained in this manner denoted as \mbox{MSE-ESTOI-DNN} offer similar improvements in ESTOI measure as observed with \mbox{MSE-DNN}  while mitigating the losses in the SDR performance to 0.2 on an average. Moreover, authors in \cite{kolbaek2018monaural} noted that MSE based systems performed at par with their proposed STOI optimization approach. We thus acknowledge the utility of MSE optimization towards the final goal of optimizing for improvements in intelligibility. It therefore makes sense to use MSE objective along with  the proposed ESTOI objective, an observation which also serves as the motivation for \textit{Case c}. 


\section{Conclusion and future work}\label{sec:conclusions}
\vspace{-1ex}
In this work, we proposed a novel objective function for optimizing objective intelligibility performance of DNN-based speech separation systems, here in terms of ESTOI, and compared it with commonly used MSE objective. We showed that the proposed approach offers improvements or performs at par with the baseline. We also showed that  a pretraining strategy utilizing weights of MSE optimized DNN as the initial point of optimization for our approach can mitigate the losses in terms SDR resulting from using ESTOI optimization alongwith preserving superior or at par intelligibility performance in terms of ESTOI. This observation alongwith results previously reported in \cite{kolbaek2018monaural}, indicate the usefulness of MSE optimization for the goal of improving intelligibility performance. The future work includes combining the MSE and ESTOI to joint objective of the form  $\mathcal{L}=\mathcal{L}_{mse}+\alpha \mathcal{L}_{estoi}$, where  $\mathcal{L}_{mse}$ and $\mathcal{L}_{estoi}$ are MSE and ESTOI losses respectively, and $\alpha$ is  a weighing parameter.


\bibliographystyle{IEEEtran}
\small

\end{document}